\documentclass[review]{elsarticle}

\usepackage{lineno,hyperref}
\usepackage{siunitx}
\modulolinenumbers[1]

\journal{arXiv}

\begin{document}

\begin{frontmatter}

\title{Half-Heusler thermoelectric materials: NMR studies}

\author[1]{Yefan Tian}
\author[1]{Nader Ghassemi}
\author[2,3]{Wuyang Ren}
\author[2]{Zhu Hangtian}
\author[2,4]{Shan Li}
\author[4]{Qian Zhang}
\author[3]{Zhiming Wang}
\author[2]{Zhifeng Ren}
\author[1]{Joseph H. Ross, Jr.\corref{correspondingauthor}}
\ead{jhross@tamu.edu}
\cortext[correspondingauthor]{Corresponding author.}
\address[1]{Department of Physics and Astronomy, Texas A\&M University, College Station, TX 77843, USA}
\address[2]{Department of Physics and Texas Center for Superconductivity at the University of Houston (TcSUH), University of Houston, Houston, TX 77204, USA}
\address[3]{Institute of Fundamental and Frontier Sciences, University of Electronic Science and Technology of China, Chengdu 610054, China}
\address[4]{Department of Materials Science and Engineering, Institute of Materials Genome \& Big Data, Harbin Institute of Technology, Shenzhen, Guangdong 518055, China}

\begin{abstract}
We report $^{59}$Co, $^{93}$Nb, and $^{121}$Sb nuclear magnetic resonance (NMR) measurements combined with density functional theory (DFT) calculations on a series of half-Heusler semiconductors, including NbCoSn, ZrCoSb, TaFeSb and NbFeSb, to better understand their electronic properties and general composition-dependent trends. These materials are of interest as potentially high efficiency thermoelectric materials. Compared to the other materials, we find that ZrCoSb tends to have a relatively large amount of local disorder, apparently antisite defects. This contributes to a small excitation gap corresponding to an impurity band near the band edge. In NbCoSn and TaFeSb, Curie-Weiss-type behavior is revealed, which indicates a small density of interacting paramagnetic defects. Very large paramagnetic chemical shifts are observed associated with a Van Vleck mechanism due to closely spaced $d$ bands splitting between the conduction and valence bands.  Meanwhile, DFT methods were generally successful in reproducing the chemical shift trend for these half-Heusler materials, and we identify an enhancement of the larger-magnitude shifts, which we connect to electron interaction effects. The general trend is connected to changes in $d$-electron hybridization across the series.
\end{abstract}


\end{frontmatter}

\section{Introduction}
In recent years, thermoelectric materials as one of the most promising solutions to the energy crisis have drawn significant attention. The thermoelectric effect is based on the direct conversion of temperature differences to electric energy or vice versa. In real-life applications, waste heat can be better collected and utilized to improve energy conversion efficiency by taking advantage of the thermoelectric effect. To optimize the performance of thermoelectric materials, the thermoelectric figure of merit $zT = \sigma S^2T/(\kappa_e+\kappa_l)$ needs to be maximized, where $S$ is Seebeck coefficient, $\sigma$ electrical conductivity, $\kappa_e$ electronic thermal conductivity and $\kappa_l$ lattice thermal conductivity, and where $\sigma S^2$ is the power factor.

Among a number of thermoelectric material families, the half-Heusler semiconductors have been of considerable interest due to their excellent thermoelectric performance. For example, NbFeSb has drawn great attention due to its relatively high earth abundance and large power factor up to 100 $\mu$W\,cm$^{-1}$\,K$^{-2}$ \cite{he2016achieving,ren2018ultrahigh}. TaFeSb has been recently reported as a stable thermoelectric candidate and shown promising thermoelectric properties \cite{zhu2019discovery}. ZrCoSb-based materials have also been widely studied and shown promising $zT$ and low thermal conductivity \cite{he2014investigating}. To better understand the mechanism and further improve thermoelectric performance, the NMR technique can be used as a local probe to reveal information about electronic properties, phonon behaviors and native defects \cite{tian2019charge,ghassemi2018structure,tian2018native}. In this work, we have applied NMR in studying members of this series of transition-metal-based half-Heusler materials, focusing in particular on trends in the electronic properties in this thermoelectric materials family.

\section{Experimental and computational details}
All half-Heusler polycrystalline samples were made from high purity elements. ZrCoSb and NbFeSb were prepared by arc-melting, ball-milling and hot-pressing and NbCoSn and TaFeSb were made by ball-milling and hot-pressing. (a) ZrCoSb is the same material described in Ref.~\cite{he2017improved} using AC hot pressing at 1400 K and pressure of $\sim$80 MPa for 2 min. Microprobe measurements indicated a composition of ZrCo$_{1.02}$Sb$_{0.99}$, while the Seebeck coefficient result indicates that ZrCoSb is an $n$-type material with a large Seebeck peak of about $-280$ $\mu$V/K observed near 600 K. (b) NbCoSn was prepared as described in Ref.~\cite{li2019n}, where the sintering was at 1113-1273 K under a pressure of $\sim$80 MPa for 2 min. The actual composition of NbCoSn was measured to be Nb$_{31.4}$Co$_{35.4}$Sn$_{33.2}$, indicating a likely $n$-type behavior of NbCoSn with extra Co possibly located on Nb sites. (c) TaFeSb was prepared as described in Ref.~\cite{zhu2019discovery} with hot-pressing temperature about 1123 K and pressure of $\sim$80 MPa for 2 min. The positive Seebeck coefficient result indicates that TaFeSb is a $p$-type material with no peak observed in $S$ below 1000 K, unlike its ``twin material'' NbFeSb \cite{tian2018native}. (d) NbFeSb is the same sample as NbFeSb-1050 described in Ref.~\cite{tian2018native}. Hall measurement showed this sample as $p$-type with carrier concentration of $9 \times 10^{19}$ cm$^{-3}$.
 
$^{59}$Co, $^{93}$Nb, and $^{121}$Sb NMR experiments were carried out by applying a custom-built pulse spectrometer at a fixed magnetic field 9 T using shift standards aqueous K$_3$[Co(CN)$_6$], NbCl$_5$, and KSbF$_6$ in acetonitrile, respectively, with positive shifts here denoting paramagnetic sign. NMR spectra were obtained from the fast Fourier transform of the spin echo using a standard $\pi/2$-$\tau$-$\pi$ sequence. 

Density function theory (DFT) calculations were conducted using the WIEN2k package \cite{blaha2020wien2k}. In this package, an (linearized) augmented plane wave plus local orbitals method is implemented. These calculations were done using the experimental lattice parameters $a = 6.068$ \si{\angstrom} for ZrCoSb \cite{marazza1975some}, 5.950 \si{\angstrom} for NbFeSb \cite{tian2018native}, 5.938 \si{\angstrom} for TaFeSb \cite{zhu2019discovery}, 5.950 \si{\angstrom} for NbCoSn \cite{li2019n,buffon2016enhancement}, and 5.883 \si{\angstrom} for TiCoSb \cite{lue2009annealing}. We used $10\times10\times10$ $k$-points and adopted the exchange correlation functional introduced by Perdew, Burke, and Ernzerhof (PBE) \cite{perdew1996generalized}. The calculations were run without spin-orbit coupling or spin polarization. Calibration of calculated $^{93}$Nb chemical shifts was done by computing shifts for YNbO$_4$ and LaNbO$_4$, and comparing to the standard reference (NbCl$_5$ in acetonitrile) as established in Ref.~\cite{papulovskiy2013theoretical}. For $^{59}$Co and $^{121}$Sb, no comparable solid-state shift standard has been established, so we report relative shifts.

\section{Results and discussion}

\subsection{ZrCoSb}

\begin{figure}
\includegraphics[width=\columnwidth]{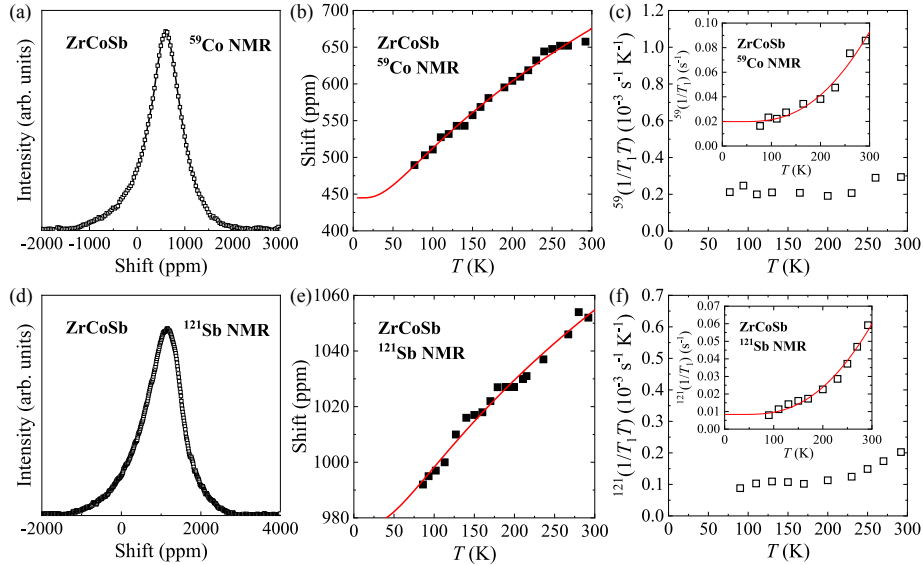}
\caption{\label{ZrCoSb} NMR results for ZrCoSb: (a) $^{59}$Co room-temperature lineshape, (b) $^{59}$Co shift vs $T$, and (c) $^{121}(1/T_1T)$ vs $T$ [inset: $^{121}(1/T_1)$ vs $T$]. (d) room-temperature $^{121}$Sb NMR lineshape, (e) $^{121}$Sb shift vs $T$, and (f) $^{59}(1/T_1T)$ vs $T$ [inset: $^{59}(1/T_1)$ vs $T$]. Shifts and $1/T_1$ for both nuclei are fitted to an excitation energy gap function (solid curves).}
\end{figure}

In Fig.~\ref{ZrCoSb}(a), the room-temperature $^{59}$Co NMR spectrum for ZrCoSb is shown. The result is relatively broad, with a full width at half maximum (FWHM) of 750 ppm (equivalent to 67 kHz), and a Lorentzian shape similar to what has been reported for TiCoSb after annealing \cite{lue2009annealing}. The corresponding ZrCoSb $^{59}$Co center-of-mass shift was measured in the temperature range 77-295 K. As shown in Fig.~\ref{ZrCoSb}(b), the results can be fitted to a function corresponding to a small energy gap. In this model, when temperature increases, the carriers are thermally excited across an excitation gap, leading to the increase of spin susceptibility. Thus, the mechanism of shift increase can be written as
\begin{equation}\label{shift}
K = K_1 + C_1T^{1/2}\exp(–\Delta/k_BT),
\end{equation}
where $K_1$ is a temperature-independent contribution and $\Delta$ is the excitation energy. Note that very similar excitation behavior was also reported in TiCoSb \cite{lue2009annealing}, with Ti substituted from the same group. The solid red curve in Fig.~\ref{ZrCoSb}(b) represents a fit to Eq.~(\ref{shift}), yielding $\Delta = E_g/2 = 18$ meV, consistent with the result obtained via spin-lattice relaxation shown below. The constant term $^{59}K_1$ is $445 \pm 20$ ppm.

Note that here we use the notation $K$ for the total observed shift of the NMR line, which is normally a sum of the Knight shift due to carrier spin susceptibility (which we designate $K_c$) and the chemical shift ($\delta$) due to local orbital susceptibility. (Although these nuclei have electric quadrupole moments, quadrupole contributions to the shift will vanish in these cubic materials.) Thus, $K = K_c + \delta$. In the limit of zero carrier density, $K_c$ will vanish, leaving only $\delta$ as the observed contribution.

$^{59}$Co spin-lattice relaxation results $^{59}(1/T_1)$ are shown in the Fig.~\ref{ZrCoSb}(c) inset, with $^{59}(1/T_1)$ shown in the main plot. Generally, the small $^{59}(1/T_1)$ indicates a relatively low density of carriers interacting with the nuclei, while at high $T$, $^{59}(1/T_1)$ rises rapidly with temperature increasing, which is the characteristic behavior for semiconductors. Similar to the shift behavior, $^{59}(1/T_1)$ was also fitted to an excitation energy gap. Based on the same mechanism as the shift, the increase in relaxation rate is due to an increase of carriers induced by thermal excitation. In this case, the spin-lattice relaxation rate is given by
\begin{equation}
1/T_1 = C_2T^2\exp(–\Delta/k_BT) + \mathrm{const.}
\end{equation}
In good agreement with the shift fitting, this fit yields an excitation gap $\Delta = 17$ meV.

Similar to $^{59}$Co NMR, $^{121}$Sb NMR has also been measured for ZrCoSb as shown in Figs.~\ref{ZrCoSb}(d)-\ref{ZrCoSb}(f). Again a broad Lorentzian-type line is observed, with room-temperature FWHM of 1030 ppm (equivalent to 93 kHz). The same fits described for $^{59}$Co have been done for the $^{121}$Sb NMR shifts and $^{121}(1/T_1T)$, with $\Delta = 10$ meV obtained from the shift fit and $\Delta = 22$ meV from $^{121}(1/T_1T)$, consistent with the $^{59}$Co NMR results and confirming the presence of a small gap. For consistency, fits for both the center-of-mass shift [Fig.~\ref{ZrCoSb}(e)] and $^{121}(1/T_1)$ [Fig.~\ref{ZrCoSb}(f) inset] have been recalculated, fixed with an excitation gap $\Delta = 17.5$ meV, the mean value obtained for $^{59}$Co. The corresponding constant term $^{121}K_1$ is 978 ppm.

In Ref.~\cite{lue2009annealing}, there is an increased line broadening observed in TiCoSb after annealing. The broad lineshape was fitted to Lorentzian function which apparently indicated dilute magnetic local-moment broadening. A quantitative analysis \cite{lue2009annealing} showed about 2\% Co local moments appearing after annealing. Similarly, the somewhat larger line widths observed in ZrCoSb also likely indicate a local moment density on this order, presumably due to Co antisites. While there has previously been uncertainty \cite{he2017improved} about the significance of antisites in this material, the relatively large line widths in ZrCoSb, as compared to other half-Heusler samples described below, point to an enhanced tendency towards such local disorder in this material.

\subsection{NbCoSn}

\begin{figure}
\includegraphics[width=\columnwidth]{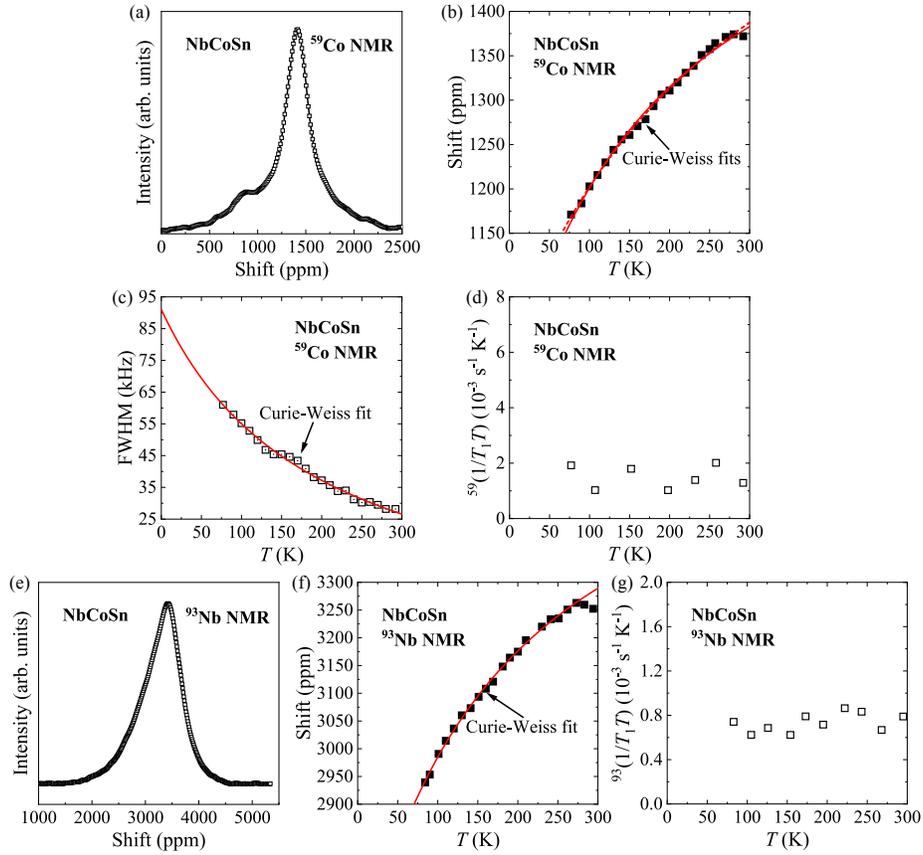}
\caption{\label{NbCoSn} NbCoSn NMR: (a)-(d) $^{59}$Co results: (a) room-temperature lineshape, (b) shift vs $T$, (c) FWHM vs $T$, and (d) $^{59}(1/T_1T)$ vs $T$. (e)-(g) $^{93}$Nb NMR results: (e) room-temperature lineshape, (f) shift vs $T$, and (g) $^{93}(1/T_1T)$ vs $T$ rates of NbCoSn. Solid and dashed curves are Curie-Weiss fits described in text.}
\end{figure}

Figs.~\ref{NbCoSn}(a) and \ref{NbCoSn}(e) show the $^{59}$Co and $^{93}$Nb room-temperature spectra of NbCoSn, respectively. The line shapes are slightly asymmetric, with line widths of 300 ppm (28 kHz) for $^{59}$Co NMR and 790 ppm (72 kHz) for $^{93}$Nb NMR, increasing as $T$ is lowered. Figs.~\ref{NbCoSn}(b) and \ref{NbCoSn}(d) show $^{59}$Co NMR center-of-mass shifts and spin-lattice relaxation rate results for NbCoSn vs $T$. Similarly Figs.~\ref{NbCoSn}(f) and \ref{NbCoSn}(g) show $^{93}$Nb NMR shifts and spin-lattice relaxation rate results. The shifts in both cases were modeled by $K = K_1 + K_2(T)$, where $K_2(T)$ was well fitted to a Curie-Weiss type function, $A_1/(T + \theta)$. The fit for $^{59}$Co $K_2(T)$ is shown as the dashed curve in Fig.~\ref{NbCoSn}(b), yielding $^{59}K_1 = 1815$ ppm and the Curie-Weiss temperature $\theta = 309$ K. Shown in Fig.~\ref{NbCoSn}(c), the $^{59}$Co line width also follows a Curie-Weiss law, FWHM $\propto A_2/(T+\theta)$, with the fitted Curie-Weiss temperature $\theta = 196$ K. In this material, the lineshapes have a tail on the low frequency side, presumably due to a small amount of composition inhomogeneity; however, the FWHM is relatively unaffected by this. Thus, we rely upon the FWHM result as a better measure of the temperature dependence, and the Curie-Weiss temperature was fixed to $\theta = 196$ K for fitting the $^{59}$Co shift, giving the result shown as the solid curve with the resulting $^{59}K_1 = 1652$ ppm.

Shown in Fig.~\ref{NbCoSn}(f), the $^{93}$Nb center-of-mass shift also follows a Curie-Weiss behavior. For both nuclei, there is a small shift downturn near room temperature, perhaps due to a small increase in carrier density due to native defects, such as seen in NbFeSb \cite{tian2018native}. The Curie-Weiss fit yields $\theta = 188$ K with the corresponding $^{93}K_1 = 3728$ ppm, in good agreement with the $^{59}$Co line width result, confirming the paramagnetic mechanism governing the observed behavior of NbCoSn. The Curie-Weiss term observed in the NbCoSn shifts indicates local hyperfine interactions with magnetic moments, as opposed to the results observed in ZrCoSb and NbFeSb (shown below), which correspond to long-range dipolar interactions with relatively dilute local moments. Thus, the behavior is that of a more strongly magnetic material, comparable to Refs.~\cite{chi2005nmr,ding2013nmr}, with the shifts and line widths scaling together. However, the effect here is much smaller than expected for a magnetic semiconductor \cite{ding2013nmr}, with for example no sign of the constant-$1/T_1$ behavior characteristic of interaction-driven fluctuations, indicating a weak effect. The source of the magnetic moments is apparently the observed Co excess, an effect which also drives this material to $n$-type behavior \cite{li2019n}.

The constant $1/T_1T$ shown in Figs.~\ref{NbCoSn}(d) and \ref{NbCoSn}(g) is characteristic of a Korringa-type mechanism, representing the partial Fermi-level DOS of the probed site. The results reveal that there are sufficient carriers so that the Fermi level locates in the band for the whole measured temperature range. The results are larger than observed in ZrCoSb, but $^{93}(1/T_1T)$ is still considerably smaller than observed in NbFeSb \cite{tian2018native}, for which the Korringa contribution was specifically identified and extracted from the paramagnetic contribution. This indicates that the Knight shift contribution is relatively small for NbCoSn.

\subsection{TaFeSb}

\begin{figure}\center
\includegraphics[width=0.7\columnwidth]{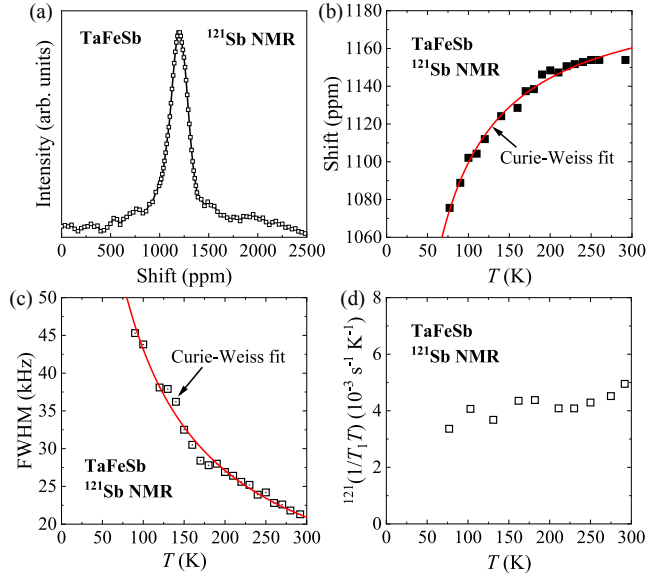}
\caption{\label{TaFeSb} $^{121}$Sb NMR for TaFeSb: (a) room-temperature lineshape, (b) shift vs $T$, (c) FWHM vs $T$, and (c) $^{121}(1/T_1T)$ vs $T$ results of TaFeSb. Solid curves are Curie-Weiss fits described in text.}
\end{figure}

Fig.~\ref{TaFeSb}(a) shows the room-temperature $^{121}$Sb NMR spectrum for TaFeSb, showing a relatively narrow FWHM equal to 210 ppm (21 kHz). Fig.~\ref{TaFeSb}(b) shows the shift obtained from the center of mass of the measured spectra. As above, the shifts can be expressed by $K = K_1 + K_2(T)$, where $K_1$ is a temperature-independent term and $K_2(T)$ is the temperature-dependent term which we observed to follow a Curie-Weiss type behavior, $A_1/(T + \theta)$. A fit to this function yields $^{121}K_1 = 1193 \pm 6$ ppm and the corresponding Curie-Weiss temperature $\theta = 9$ K. The fitting shows the existence of paramagnetic centers, possibly in the form of paramagnetic defects; however, the magnitude is considerably smaller than what was observed in NbCoSn.

Fig.~\ref{TaFeSb}(c) shows the temperature dependence of the $^{121}$Sb FWHM, which can also be fitted to a Curie-Weiss type function FWHM $\propto A_2/(T+\theta)$. The unconstrained fit yields $\theta = 22$ K, close to the $^{121}$Sb shift result. For the same reason as NbCoSn, we thus fixed the Curie-Weiss temperature as 22 K for the shifts with corresponding $^{121}K_1 = 1199$ ppm, yielding the fitting curve shown in Fig.~\ref{TaFeSb}(b). (In the figure, the $\theta = 9$ and 22 K curves are indistinguishable, so the $\theta = 9$ K case cannot be seen.) Similar to Refs.~\cite{chi2005nmr,ding2013nmr}, and for NbCoSn, the FWHM of TaFeSb at each temperature also scales as about 2-3 times the shift, thus the average local field is about the same size as the distribution of local fields about the mean value. This confirms the local-hyperfine mechanism, and Curie-Weiss behavior of TaFeSb. By analogy to the previous analysis of the ``twin material'' NbFeSb \cite{tian2018native}, the local moments for TaFeSb are likely Fe antisites. The Curie-Weiss behavior points to weak interactions for these moments, in contrast to NbFeSb (below) where the $1/T_1$ peak indicates independent fluctuations of dilute moments, behavior which is not observed in TaFeSb.

The constant $1/T_1T$ shown in Fig.~\ref{TaFeSb}(d) represents the partial Fermi-level DOS of the probed site, indicating the weakly metallic behavior of TaFeSb at low temperature. When temperature is above 250 K, a small upturn can be observed, revealing a possible additional excitation of carriers, as was fitted for the case of ZrCoSb. However, the baseline $1/T_1T$ in this case is larger, making the carrier excitation effect relatively quite small. The constant shift term $K_1$ is a combination of chemical shift and Knight shift. Based on the Korringa relation, the Knight shift $^{121}K_c$ can be estimated from measured $^{121}(1/T_1T)$ and the small magnitude reveals that the contribution of charge carrier is not significant. Details of calculation can be found below.

\subsection{NbFeSb}

\begin{figure}
\includegraphics[width=\columnwidth]{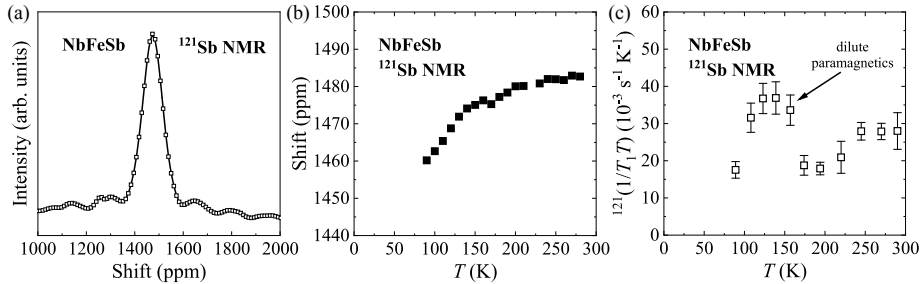}
\caption{\label{NbFeSb} $^{121}$Sb NMR: (a) room-temperature lineshape, (b) shift vs $T$, and (c) $^{121}(1/T_1T)$ vs $T$ results of NbFeSb.}
\end{figure}

Figs.~\ref{NbFeSb}(a)-\ref{NbFeSb}(c) show the $^{121}$Sb NMR results for NbFeSb, which shows similar behavior comparable to $^{93}$Nb NMR reported in Ref.~\cite{tian2018native} for the same sample. In Fig.~\ref{NbFeSb}(a), the $^{121}$Sb NMR spectrum shows a very narrow line width, with FWHM about 90 ppm (8 kHz) at room temperature, similar to the $^{93}$Nb results. There is a very small decrease in shift as temperature is lowered [Fig.~\ref{NbFeSb}(b)], and as shown in Fig.~\ref{NbFeSb}(c), $^{121}(1/T_1T)$ also shows a peak very similar to the $^{93}$Nb results \cite{tian2018native}, shown to be due to very dilute magnetic moments \cite{tian2018native}. To extract the moment density, it requires a more detailed set of measurements using a stretched-exponential analysis \cite{tian2018native}; however, for comparison the observed peak $^{93}(1/T_1)$ corresponds to $^{93}(1/T_1T) = 30 \times 10^{-3}$ (s\,K)$^{-1}$, very similar to the result observed here. Due to a long range dipole mechanism, in the dilute-moment limit, the results are scaled only by the squared gyromagnetic ratios of the two nuclei \cite{he2014investigating}, which are almost identical. Both $^{93}$Nb and $^{121}$Sb relaxation peaks have very similar magnitudes further confirming the mechanism of long-range dipolar interaction with dilute local magnetic spins in NbFeSb and validating the previous results.

NbFeSb has a significant Knight shift for both nuclei. In Ref.~\cite{tian2019defect}, both $^{93}$Nb and $^{121}$Sb chemical shifts were extracted by studying a series of Ti-substituted samples (Nb$_{1-x}$Ti$_x$FeSb). Assuming that the chemical shift is linearly dependent on Ti-substitution level for these semiconductors, the chemical shifts were then obtained by a fitting model with measured shifts composed of carrier-density-dependent Knight shift and substitution-fraction-dependent chemical shift. As a result, it also turned out that both $^{93}$Nb and $^{121}$Sb also have significant Knight shifts in the undoped material. This is consistent with the measured much larger $1/T_1T$ compared to other half-Heusler materials: for example, underlying the paramagnetic peak [Fig.~\ref{NbFeSb}(c)] it can be seen that the baseline $1/T_1T$ is on order of $^{121}(1/T_1T) \approx 15 \times 10^{-3}$ (s\,K)$^{-1}$.

\subsection{DFT computed shifts}

\begin{figure}\center
\includegraphics[width=0.6\columnwidth]{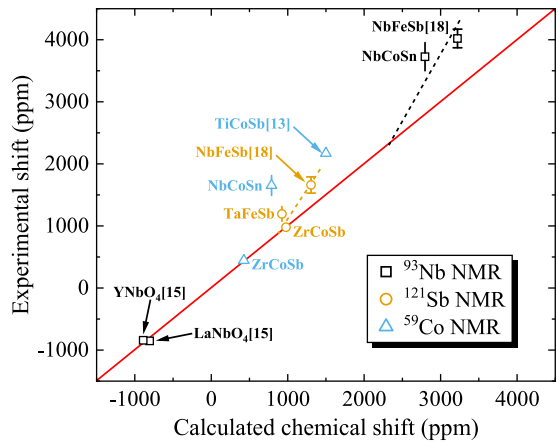}
\caption{\label{DFTvsEXP_shifts} Comparison between experimental shifts and DFT calculated chemical shifts. NbFeSb chemical shifts are from Ref.~\cite{tian2019defect}, with error bars based on the subtraction of Knight shift contribution as described there, TiCoSb shift from Ref.~\cite{lue2009annealing} and Y(La)NbO$_4$ from Ref.~\cite{papulovskiy2013theoretical}. Vertical bars without caps for the other compositions show the estimated range of chemical shifts after subtraction of Knight shift based on Korringa relation. Additional errors corresponding to uncertainty in measured shift values are smaller than the visible symbols. Dashed lines are guides to the eye.}
\end{figure}

Fig.~\ref{DFTvsEXP_shifts} is the comparison between experimental shifts and DFT calculated chemical shifts ($\delta$) for the half-Heusler materials, including results for all nuclei ($^{93}$Nb, $^{59}$Co and $^{121}$Sb) measured in this work. The experimental values plotted are the temperature-independent shifts ($K_1$), extracted as discussed above, from which the effects of thermally excited carriers (ZrCoSb) or Curie-Weiss contributions (NbCoSn, TaFeSb) were removed. While these values generally contain Knight shifts and chemical shifts ($K = K_c +\delta$), where possible, we have identified the Knight shift contributions, as discussed below. We also include TiCoSb \cite{lue2009annealing}, another member of this general family for which NMR studies have been reported, and for NbFeSb the values are fitted values of $\delta$ alone from composition-dependent results. The red line in Fig.~\ref{DFTvsEXP_shifts} corresponds to perfect agreement between experiment and calculation. Since measured chemical shifts are generally reported relative to NMR shift standards having unknown absolute shift contributions, a computational reference can be used to calibrate the absolute DFT shifts. For $^{93}$Nb NMR shift, the calibrating references are YNbO$_4$ and LaNbO$_4$. For $^{59}$Co and $^{121}$Sb for which no comparable materials with well-defined chemical shift are available, we discuss here the relative shifts.

The Korringa relation is an important tool to understand metallic behavior, which is characterized by a $T$-independent Knight shift ($K_c$) and a constant $1/T_1T$. Often \cite{carter1977metallic}, the product $K_c^2T_1T$ is found to be very close to $K_c^2T_1T = (\hbar\gamma_e^2)/(4\pi k_B\gamma_n^2)$,where $\gamma_e$ and $\gamma_n$ are the gyromagnetic ratios of the electron and nucleus. However, since other mechanisms can contribute to $1/T_1$ and generally increase the rate, here we can consider that $1/T_1T$ serves to provide an upper limit estimate for $K_c$. For example, in substituted NbFeSb a significant orbital contribution was shown to contribute to the relaxation rate \cite{tian2019defect}; however, this term does not contribute to $K_c$ \cite{tian2019defect}, and as noted above, paramagnetic contributions due to local moments can in some cases also contribute to $1/T_1$. For $^{121}$Sb, based on the standard product $K_c^2T_1T = 4.5 \times 10^{-6}$ s\,K, the low temperature $^{121}(1/T_1T)$ values correspond to 20 ppm for ZrCoSb, and 130 ppm for TaFeSb. The small contribution for ZrCoSb, and a similar small contribution estimated from the $^{59}$Co relaxation rates, justifies putting those values on the red line in Fig.~\ref{DFTvsEXP_shifts}, and using these values to calibrate the relative $^{59}$Co and $^{121}$Sb chemical shifts. For NbCoSn, the mean observed $^{93}(1/T_1T) = 1.5 \times 10^{-3}$ (s\,K)$^{-1}$ is 10 times smaller than the Korringa $^{93}(1/T_1T)$ extracted for the NbFeSb sample studied here \cite{he2014investigating}, hence we estimate that $^{93}K_c$ may be $\sqrt{10}$ times smaller than the extracted Knight shift for NbFeSb, or about 230 ppm. The $^{59}(1/T_1T)$ values for NbCoSn are somewhat smaller than for $^{93}$Nb, but $K_c$ in both cases is expected to be negative, due to the negative core-polarization hyperfine fields corresponding to the $d$ states which dominate the electronic behavior at the band edges for these ions. We have placed vertical bars on the NbCoSn points in Fig.~\ref{DFTvsEXP_shifts}, to represent the corresponding $K_c$ contributions which may be present, and hence showing the range of underlying chemical shifts which can be deduced from the experimental results. These estimated uncertainties have been included in Fig.~\ref{DFTvsEXP_shifts} as simple vertical lines. For the case of NbFeSb, as described above measurements of Ti-substituted materials were previously used to extract the chemical shifts \cite{tian2019defect}, and the corresponding statistical error bars from this fit are displayed in Fig.~\ref{DFTvsEXP_shifts}.

\subsection{Discussion and analysis}
The actual size of the bulk gap in ZrCoSb is believed to be much larger than the fitting result described above (for example a calculated value near 1 eV has been reported \cite{lee2011electronic,zahedifar2018band}, and the observed Seebeck coefficient peak \cite{he2017improved} for this sample also points to a gap that is significantly larger than what is apparent from the NMR results), indicating that a defect level locates above the valence band forming an impurity band in the bulk gap, similar to NbFeSb \cite{tian2018native}. In Ref.~\cite{tian2018native}, NbFeSb has also exhibited a very small gap $\sim$30 meV, shown to indicate an impurity band rather than a real band gap. By analogy, this small gap for ZrCoSb is likely to indicate the existence of an impurity band. The $n$-type carrier suggests that the impurity band is right below the conduction band, different from NbFeSb. Note also that $1/T_1T$ for both nuclei measured in ZrCoSb is very small, especially at low temperatures. As discussed below, it can be concluded that the extracted $K_1$ values have very little contributions due to free carriers (hence small Knight shifts, $K_c$). Thus, we analyze the fitted $K_1$ for ZrCoSb as representing the chemical shift. A similar argument can be applied to the previously reported results for TiCoSb \cite{lue2009annealing}, as further discussed below.

Comparing to ZrCoSb, the narrow line width of NbFeSb indicates a particularly strong tendency for local ordering in NbFeSb, for which the observed $1/T_1T$ peak provides evidence of a small density of independent magnetic moments. On the other hand, both TaFeSb and NbCoSn exhibit Curie-Weiss behavior, which can be understood as due to interacting Fe or Co antisites, owing to stoichiometry differences. The off-stoichiometry for TaFeSb is probably driven by difficult synthesis conditions due to the high melting temperature of Ta, while compared with TaFeSb, the broader line width of NbCoSn corresponds to its larger tendency for off-stoichiometry, and is reasonably consistent with the 2\% Co excess measured by EDS \cite{li2019n}.

The DFT computed shifts capture the general trend of the experimental shifts rather well, as seen in Fig.~\ref{DFTvsEXP_shifts}. Although there are Knight shifts involved in some of the measured shifts, as discussed above their estimated magnitudes are relatively small compared to the very large chemical shifts. This gives further confidence that WIEN2k can provide good predictions of chemical shift for these half-Heusler materials, and that the very large range of observed shifts is indeed intrinsic to these materials. For ZrCoSb, we showed that both the $^{59}$Co and $^{121}$Sb NMR measured shifts include very small Knight shifts, with a Korringa-type contribution that is essentially negligible. Similarly, TaFeSb also falls close to the red line in Fig.~\ref{DFTvsEXP_shifts}. However, a trend can be seen by which the larger shifts are enhanced relative to the expected values; this is emphasized by the dashed lines in Fig.~\ref{DFTvsEXP_shifts}. Note also that these larger shift values ($^{93}$Nb shift for NbFeSb and NbCoSn; $^{121}$Sb shift for NbFeSb and TiCoSb) are out of the established ranges of chemical shifts for $^{93}$Nb \cite{mason1989nmr} and $^{121}$Sb \cite{takashima1995theoretical} NMR.

These paramagnetic shifts can be understood as due to the paramagnetic susceptibility of nearly degenerate $d$ bands split into the conduction and valence bands. The mechanism is closely related to the Van Vleck susceptibility, $\chi_\mathrm{VV}$, which measures the macroscopic orbital paramagnetism. A general form of the NMR orbital shift can be expressed \cite{slichter2013principles} as
\begin{equation}
K_\mathrm{orb}=\frac{2e^2}{m^2 c^2}\sum\frac{\langle \Psi | L_z | \Psi' \rangle  \langle \Psi' | L_z/r^3 | \Psi \rangle}{\Delta E}+\mathrm{c.c.},
\end{equation}
where $\Psi$ is an occupied state, $\Psi'$ is an excited state, and the sum of angular momentum matrix elements across the semiconducting gap is an integration throughout momentum space of vertical transitions, covering associated with the bands as a whole rather than being dominated by the band edges. The multiplicity and larger angular momentum of $d$ states split across the gap, along with a relatively small $\Delta E$ \cite{slichter2013principles}, can lead to a large orbital shift contribution. Since the sum is unaffected by small additions of charger carriers at the band edge, this shift also has little temperature dependence.

As noted above, in Fig.~\ref{DFTvsEXP_shifts} it can been observed that the experimental shift has the trend of exceeding the calculated chemical shift, as the shift values become large. This is seen most clearly for $^{121}$Sb and $^{93}$Nb shifts, illustrated by dashed lines. A likely explanation for this behavior is described in Refs.~\cite{kontani1996magnetic,kontani1997theory,udagawa2010interplay}, indicating that the Van Vleck susceptibility can be enhanced by electron-electron interactions as well as the more familiar enhancement of the Pauli susceptibility. There is also an increase due to spin-orbit interactions, with the example \cite{udagawa2010interplay} being Sr$_2$RuO$_4$, for which the spin-orbit strength should be comparable in magnitude to NbFeSb, and the effect relies upon inter-orbital interactions, and thus requires orbital degeneracy within the band. Thus, tentatively we ascribe the orbital dominated Van Vleck shift enhancement in the half-Heuslers to $e$-$e$ interactions in nearly degenerate $d$ orbitals.

\begin{figure}\center
\includegraphics[width=0.6\columnwidth]{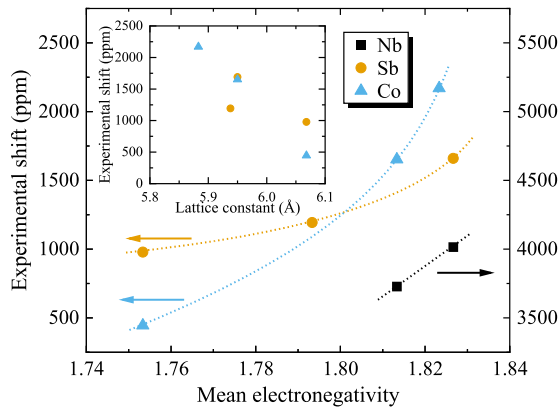}
\caption{\label{electronegativity} Experimental shifts vs mean electronegativity. Mean electronegativity is the mean value of Pauli electronegativities of three elements in each ternary half-Heusler sample. Dashed curves are guides to the eye. Inset: Co and Sb shifts vs half-Heusler lattice constant.}
\end{figure}

The correlation between chemical shifts and the electronegativities of neighboring ions is a comparison that is often made in NMR spectroscopy. In a study of main-group half-Heusler materials \cite{dupke2014systematic}, it was also recently shown that the metal-atom chemical shifts increase nearly linearly with the anion electronegativity. Such a comparison between anions cannot be made for the transition-metal half-Heusler materials studied here, since most of the corresponding compounds containing other pnictogen or carbon-group anions are not stable in the same structure. However, a regular trend can be seen by comparing the chemical shifts to the mean Pauli electronegativities of the 3 elements comprising each compound, as shown in Fig.~\ref{electronegativity}.

The general trend displayed in Fig.~\ref{electronegativity} is an increase in shift vs mean electronegativity. Since in each series the atom corresponding to the nucleus being measured is held constant, increasing mean electronegativity implies that the observed nucleus experiences a nominal decrease in on-site charge. Thus, this is the same trend observed vs anion substitution in main-group half-Heusler compounds \cite{dupke2014systematic}. However, whereas in proton NMR a decrease in on-site charge will increase shift because the electron density contributes most significantly to the diamagnetic NMR shielding, here the large positive NMR shifts seen here are clearly paramagnetic shifts dominated by a Van Vleck mechanism as discussed above. The results are not completely linear, in fact for $^{59}$Co and $^{121}$Sb the upward trend observed for large shifts is the same as the trend identified in comparison to the DFT results discussed here, with large chemical shifts becoming enhanced in magnitude over what is expected.

Along with the electronegativity trend, there is generally a corresponding decrease in shift vs increasing lattice constant, $a$. (The $^{93}$Nb shifts are not plotted since they are for compounds with the same lattice constant.) This trend can be connected to changes in hybridization, which will increase as $a$ decreases, and contribute a greater mix of $d$ orbitals in the conduction and valence bands, which as noted above produces the large observed paramagnetic chemical shifts by the Van Vleck mechanism. This also helps to understand the electronegativity trend, since moving to the right and upward on the periodic table, in the direction of increasing electronegativity, also corresponds to more compact orbitals and enhanced hybridization. Thus, the observed experimental trends and large NMR shifts can be understood as relating to an enhanced mix of orbitals in these compounds.

To investigate whether this might be due to effects such as stoichiometry or site disorder, we performed a volume minimization for these materials, using the PBE functional in WIEN2k, and obtained $5.948 \pm 0.004$ \si{\angstrom} for TaFeSb and $5.961 \pm 0.003$ \si{\angstrom} for NbFeSb. The calculated lattice constant difference between TaFeSb and NbFeSb is in a good agreement with the experimental difference, and it is also similar to other recently reported calculated values \cite{naydenov2019huge} indicating that the smaller size of TaFeSb lattice is an intrinsic feature. As noted above the significantly larger paramagnetic chemical shift observed in NbFeSb relative to TaFeSb is in agreement with DFT results based on the experimental results. With these large shifts dominated by hybridization among $d$ orbitals, the TaFeSb shift is smaller despite the more diffuse 5$d$ orbitals which might be expected to lead to enhanced hybridization. This result is likely caused by the presence of the 4$f$ electrons on Ta, and the resulting contraction and energetic favoring of the Ta 6$s$ over the 5$d$ electrons, thus reducing the number of occupied $d$ orbitals which can interact with the applied field.

\section{Summary}
In this work, we have investigated various half-Heusler thermoelectric materials (ZrCoSb, NbCoSn, TaFeSb and NbFeSb) using NMR as a local probe combined with DFT calculations. For ZrCoSb, both $^{59}$Co and $^{121}$Sb shift and spin-lattice relaxation measurements show consistent results indicating the excitation of carriers and the existence of impurity band right below the conduction band. For NbCoSn and TaFeSb, both show Curie-Weiss-like behavior revealing paramagnetic-type defects. The constant spin-lattice relaxation rates represent the partial Fermi-level DOS of the probed site, indicating the metallic behavior at measured temperature range. $^{121}$Sb NMR spin-lattice relaxation result for NbFeSb shows a clear peak due to long range dipolar interaction with local magnetic defects, confirming previous published $^{93}$Nb NMR results for NbFeSb. These paramagnetic shifts can be understood as degenerate $d$ bands splitting and mixing in the conduction and valence bands. The observed trends of chemical shift vs electronegativity and lattice constant can be connected to variations in the $d$-electron hybridization in half-Heuslers. The DFT computed results give an overall reasonable prediction of NMR chemical shifts for half-Heusler materials. The largest shifts are observed to exceed what is predicted, and we discuss a likely mechanism due to electron-electron enhanced Van Vleck susceptibility.

\section*{Acknowledgments}
This work done at Texas A\&M University is supported by the financial support of the College of Science Strategic Transformative Research Program (COS-STRP) at Texas A\&M University and the Robert A. Welch Foundation, Grant No. A-1526. Z.W. acknowledges the financial support from the National Key Research and Development Program of China (2019YFB2203400). Q.Z. acknowledges the financial support from the National Natural Science Foundation of China (11674078) and the Natural Science Foundation for Distinguished Young Scholars of Guangdong Province of China (2020B1515020023). 

\bibliography{mybibfile}

\end{document}